\documentclass{PoS}

\usepackage{amsmath}
 \usepackage{comment}

\title{Tensor renormalization group analysis of ${\rm CP}(N-1)$ model in two dimensions}

\ShortTitle{Tensor renormalization group analysis of CP($N-1$) model in two dimensions}

\author{\speaker{Hikaru Kawauchi}\\
        Institute for Theoretical Physics, Kanazawa University, Kanazawa 920-1192, Japan\\
        E-mail: \email{kawauchi@hep.s.kanazawa-u.ac.jp}}

\author{Shinji Takeda\\
        Institute for Theoretical Physics, Kanazawa University, Kanazawa 920-1192, Japan\\
        E-mail: \email{takeda@hep.s.kanazawa-u.ac.jp}}

\abstract{We apply the higher order tensor renormalization group to lattice CP($N-1$) model 
in two dimensions. A tensor network representation of CP($N-1$) model is derived. We confirm that 
the numerical results of the CP(1) model without the $\theta$-term using this method are consistent with that of the O(3) model which is analyzed by the same method in the region $\beta \gg 1$ and that obtained by Monte Carlo simulation in a wider range of $\beta$.}

\FullConference{The 33rd International Symposium on Lattice Field Theory\\
		14 -18 July 2015\\
		Kobe International Conference Center, Kobe, Japan}

\begin{document}

\section{Introduction}
The sign problem is known to be one of the most difficult problems in lattice gauge theory, for example, QCD with the quark chemical potential $\mu$ or including the $\theta$ term where the corresponding Boltzmann weight is complex. So far, a considerable number of research within a framework of Monte Carlo method has been devoted to overcome the sign problem, and there are limited successes, depending on dimensionality or a property of specific model. On the other hand, another possibility to avoid the sign problem is just to leave Monte Carlo method.

The tensor renormalization group (TRG) method is one of such a possibility and has no sign problem, which was originally proposed for studying two-dimensional classical systems by Levin and Nave\cite{Michael}. This method no longer regards the Boltzmann weight as a probability of generating field configurations as in Monte Carlo method. 
The TRG method consists of two main steps. The first step is to obtain the tensor network representation of the partition function of a system. In order to obtain the representation, one has to expand the Boltzmann weight using new integers and integrate out the old degrees of freedom. The new integers will become the indices of the tensor. The next step is to reduce the number of the tensors under controlling systematic errors. After the number of the coarse grained tensors decreases, it is possible to calculate the partition function by contracting all indices of the tensors. Although the original TRG was invented for two dimensional system,
the higher order TRG (HOTRG) was introduced by Xie {\it et al.}\cite{Xie}
as an extension to higher dimensional systems.

The strong CP problem is one of the interesting topics in QCD;
why the parameter $\theta$ for CP odd operator in the QCD Lagrangian, where such a term is allowed to exist, is so small.
In order to answer the question, understanding of the non-perturbative QCD dynamics including $\theta$-term is indispensable,
but the presence of the this term causes the sign problem.
Instead of dealing with QCD directly, it is reasonable to start to investigate its toy model, CP($N-1$) model, which shares many features with QCD.
A long time ago, Schierholz suggested an interesting scenario to solve the strong CP problem in CP($N-1$) model by analyzing phase diagram in the $\beta - \theta$ plane \cite{Schierholz}.
Although it is not clear that the solution can be directly applied to QCD,
it is interesting to verify the scenario with another method, namely TRG approach which is absent of the sign problem.

In this report, we apply the HOTRG to CP($N-1$) model in two dimensions, and
present the tensor network representation and numerical results.
Although including $\theta$-term in the tensor network representation is straightforward\footnote{An explicit form of tensor in the presence of $\theta$-term shall be given in a separate paper.}, we present the tensor at $\theta=0$.

\section{Tensor network representation of CP($N-1$) model}

The partition function of lattice CP($N-1$) model is given by
\begin{align}
Z=
\int \prod_{i} dz_i dz_i^\ast \prod_{<i,j>}dU_{i,j} 
{\rm exp}\left\{
\beta N\sum_{i,j}
\Bigl[
z^\ast_i\cdot z_jU_{i,j}+
z^\ast_j \cdot z_i U_{i,j}^\dag
\Bigr]
\right\},
\end{align}
where $z_i$ is $N$-component complex scalar field of unit length, $|z|=1$, and $U_{i,j}$ is link variable described by auxiliary vector field $A_{i,j}$, i.e. $U_{i,j}={\rm exp}\{iA_{i,j}\}$.
In order to obtain a tensor network representation, one has to expand the Boltzmann weight with new integers, and then integrate out the old degrees of freedom (The complex fields $z_i$ and the auxiliary field $A$ in this case). In the end, one can obtain a tensor which has indices of the new integers.

To expand the Boltzmann weight with new integers, we use the characterlike expansion \cite{Plefka},
\begin{align}
{\rm exp}\left\{
\beta N
\Bigl[
z^\ast_i\cdot z_j U_{i,j}+
z_i \cdot z^\ast_j U_{i,j}^\dag
\Bigr]
\right\}
=
Z_0(\beta)\sum_{l,m=0}^\infty d_{(l;m)}{\rm exp}[i(m-l)A_{i,j}]h_{(l;m)}(\beta)
f_{(l;m)}(z_i,z_j), 
\label{charactercpn}
\end{align}
where $d_{(l;m)}$ are dimensionalities of characterlike representations, $h_{(l;m)}(\beta)$ are characterlike expansion coefficients, $f_{(l;m)}(z_i,z_j)$ are characterlike expansion characters, and $Z_0(\beta)$ is the normalization factor which makes $h_{(0;0)}(\beta)=1$. The integers $l$ and $m$ will become the indices of the tensor shown below.

The characterlike expansion coefficients $h_{(l;m)}(\beta)$ are expressed by the modified Bessel functions of the first kind
\begin{align}
h_{(l;m)}(\beta)=\frac{I_{N-1+l+m}(2N\beta)}{I_{N-1}(2N\beta)}.
\end{align}
Since the modified Bessel function of the first kind, $I_n(x)$, decreases rapidly as $n$ increases with a fixed value of $x$, one can safely truncate the sum of $l$ and $m$ in eq. (\ref{charactercpn}) at some order (say $l_{\rm max}$).

We show some explicit form of the dimensionalities of characterlike representations $d_{(l;m)}$ and the characterlike expansion characters $f_{(l;m)}(z_i,z_j)$ with any values of $l$.

\noindent
For $m=0$,

\begin{align}
d_{(l;0)}=\sqrt{\frac{(N-1+l)!}{l!(N-1)!}},
\end{align}
\begin{align}
f_{(l;0)}(z_i,z_j)=\sqrt{\frac{(N-1+l)!}{l!(N-1)!}}(z_i \cdot z_j^\ast)^l.
\label{fl0}
\end{align}

\noindent
For $m=1$,
\begin{align}
d_{(l;1)}
=\sqrt{\frac{(N+l)!(N-1+l)}{l!(N-1)!(N-1)}}\frac{N-1}{N-1+l},
\end{align}
\begin{align}
f_{(l;1)}(z_i,z_j)=\sqrt{\frac{(N+l)!(N-1+l)}{l!(N-1)!(N-1)}}\Bigl[(z_i\cdot z_j^\ast)^l(z_i^\ast \cdot z_j)
-\frac{l}{N-1+l}(z_i \cdot z_j^\ast)^{l-1}\Bigr].
\label{fl1}
\end{align}

\noindent
For $m=2$,
\begin{align}
d_{(l;2)}=\sqrt{\frac{(N-1+l)(N+l)(N+1+l)!}{2(N-1)l!N!}}\frac{N(N-1)}{(N-1+l)(N+l)},
\end{align}
\begin{align}
\nonumber
f_{(l;2)}(z_i,z_j)
&=\sqrt{\frac{(N+1+l)!(N-1+l)(N+l)}{2l!N!(N-1)}}\\
&\ \ \ \times \Bigl[(z_i \cdot z_j^\ast)^l(z_i^\ast \cdot z_j)^2-\frac{2l}{N+l}(z_i\cdot z_j^\ast)^{l-1}(z_i^\ast \cdot z_j)+\frac{(l-1)l}{(N-1+l)(N+l)}(z_i\cdot z_j^\ast)^{l-2}\Bigr].
\label{fl2}
\end{align}


\begin{figure}[t]
 \centering
  \includegraphics[width=50mm]{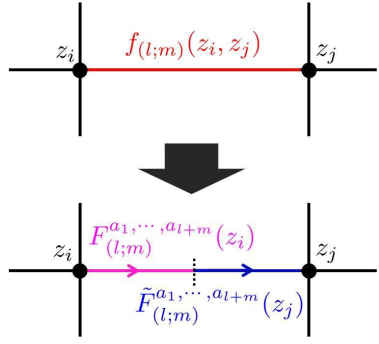}
 \caption{Decomposition of characterlike expansion characters $f_{(l;m)}(z_i,z_j)$.}
 \label{cpndecompose}
\end{figure}

The term, $f_{(l;m)}(z_i,z_j)$, is expressed by the combination of two complex scalar fields, $z_i$ and $z_j$. In order to obtain a tensor network representation, one has to integrate out the complex scalar fields $z$ site by site.
For that purpose, it is convenient to rewrite it as follows,
\begin{align}
f_{(l;m)}(z_i,z_j)
=
\sum_{\{a\}}
F^{a_1,\cdots,a_{l+m}}_{(l;m)}(z_i)
\tilde{F}^{a_1,\cdots,a_{l+m}}_{(l;m)}(z_j),
\end{align}
where $\{a\}=a_1,a_2,\cdots, a_{l+m}$, and
$a_n=1,2,\cdots, N$ for
$n=1, 2, \cdots, l+m$.
A pictorial expression of this decomposition is illustrated in Figure \ref{cpndecompose}.
The explicit forms of $F$ and $\tilde F$ are as follows.

\noindent
For $m=0$, 
\begin{align}
F^{a_1,\cdots,a_{l}}_{(l;0)}(z_i)
=
\left(\frac{(N-1+l)!}{l!(N-1)!}\right)^\frac{1}{4}z_i^{a_1}\cdots z_i^{a_l},
\end{align}
\begin{align}
\tilde{F}^{a_1,\cdots,a_{l}}_{(l;0)}(z_i)
=
\left(\frac{(N-1+l)!}{l!(N-1)!}\right)^\frac{1}{4}z_i^{\ast a_1}\cdots z_i^{\ast a_l}.
\end{align}

\noindent
For $m=1$,
\begin{align}
F^{a_1,\cdots,a_{l},a'_1}_{(l;1)}(z_i)
=
\left(\frac{(N+l)!(N-1+l)}{l!(N-1)!(N-1)}\right)^\frac{1}{4}
\Big[z_i^{a_1} z_i^{\ast a'_1} +\sqrt{\frac{l}{N(N-1+l)}}\delta^{a_1 a'_1}\Big]
z_i^{a_2} \cdots z_i^{a_l},
\end{align}
\begin{align}
\tilde{F}^{a_1,\cdots,a_{l},a'_1}_{(l;1)}(z_i)
=
\left(\frac{(N+l)!(N-1+l)}{l!(N-1)!(N-1)}\right)^\frac{1}{4}
\Big[z_i^{\ast a_1} z_i^{a'_1} -\sqrt{\frac{l}{N(N-1+l)}}\delta^{a_1 a'_1}\Big]
z_i^{\ast a_2}\cdots z_i^{\ast a_l}.
\end{align}


\noindent
For $m=2$,
\begin{align}
\nonumber
&F^{a_1,\cdots,a_{l},a'_1,a'_2}_{(l;2)}(z_i)\\
&= C_{(l;2)}
\Big[z_i^{a_1}z_i^{a_2} z_i^{\ast a'_1} z_i^{\ast a'_2}
-\frac{l(N-1+l)+\sqrt{lN(N-1+l)}}{(N-1+l)(N+l)}\delta^{a_1 a'_1}
z_i^{a_2}z_i^{\ast a'_2}\Big]
z_i^{a_3}\cdots z_i^{a_l},
\end{align}
\begin{align}
\nonumber
&\tilde{F}^{a_1,\cdots,a_{l},a'_1,a'_2}_{(l;2)}(z_i)\\
&= C_{(l;2)}
\Big[ z_i^{\ast a_1} z_i^{\ast a_2} z_i^{a'_1} z_i^{a'_2}
-\frac{l(N-1+l)-\sqrt{lN(N-1+l)}}{(N-1+l)(N+l)}\delta^{a_2 a'_2}
z_i^{\ast a_1}z_i^{a'_1}\Big]z_i^{\ast a_3}\cdots z_i^{\ast a_l},
\end{align}
with
\begin{align}
C_{(l;2)}=
\left(\frac{(N+1+l)!(N-1+l)(N+l)}{2l!N!(N-1)}\right)^\frac{1}{4}.
\end{align}


\noindent

\begin{figure}[t]
 \centering
  \includegraphics[width=120mm]{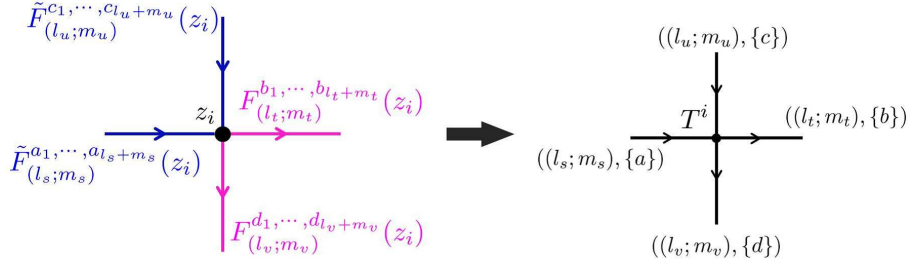}
 \caption{Integrating out the N-component complex scalar field $z_i$.}
 \label{integrateout}
\end{figure}

After the decomposition, the last step is to integrate out the old degrees of freedom, $z$ and $A$.
If we focus on a site $i$, there are two $F$ and two $\tilde{F}$, as illustrated in Figure \ref{integrateout}. 
A tensor expressed in terms of them is given by
\begin{align}
\nonumber
&T_{((l_s;m_s),\{a\})((l_t;m_t),\{b\})((l_u;m_u),\{c\})((l_v;m_v),\{d\})}\\
\nonumber
=&\int dz_i dz^\ast_i
\sqrt{d_{(l_s;m_s)}d_{(l_t;m_t)}d_{(l_u;m_u)}d_{(l_v;m_v)}
h_{(l_s;m_s)}(\beta)h_{(l_t;m_t)}(\beta)
h_{(l_u;m_u)}(\beta)h_{(l_v;m_v)}(\beta)}\\
&\hspace{50pt}
\times
\tilde{F}^{a_1,\cdots,a_{l_s}}_{(l_s;m_s)}(z_i)
F^{b_1,\cdots,b_{l_t}}_{(l_t;m_t)}(z_i)
\tilde{F}^{c_1,\cdots,c_{l_u}}_{(l_u;m_u)}(z_i)
F^{d_1,\cdots,d_{l_v}}_{(l_v;m_v)}(z_i).
\label{cpntensor}
\end{align}
The integration of the complex scalar fields, $z_i$ and $z_i^\ast$, can be done analytically and then the elements of the tensor are fixed.

At this point, we mention the integration of the link variable.
In contrast to the case of the complex scalar fields, the integration of the link variable is rather simple.
\begin{align}
\int_{-\pi}^{\pi} dA \ 
{\rm exp}\{{i(m-l)A}\}
=
\delta_{l, m}.
\end{align}
This just gives a constraint that the integer $l$ is equivalent to the integer $m$.
By using the tensor in eq.(\ref{cpntensor}), we apply the HOTRG and obtain the partition function of CP($N-1$) model.

\section{Numerical results}

\begin{figure}[t]
 \centering
  \includegraphics[width=12cm]{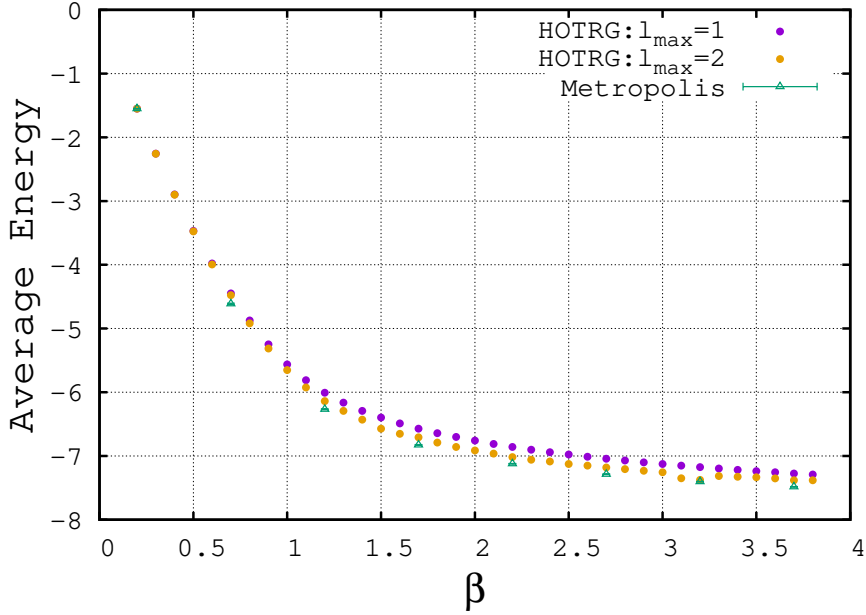}
 \caption{Average energy of CP(1) model computed by HOTRG and Metropolis algorithm. The lattice size is $4 \times 4$. The orange marks indicate the results of HOTRG and the green marks indicate the results of Metropolis algorithm. }
 \label{hotrgvsmetro}
\end{figure}

First, we compare the result of TRG method ($l_{\rm max}=1, 2$) with that of Monte Carlo simulation. Figure \ref{hotrgvsmetro} compares the average energy of CP(1) model computed by the two methods. The result of TRG method ($l_{\rm max}=2$) is almost consistent with that of Monte Carlo simulation. The little difference between the two results is considered to the truncation error $l_{\rm max}=2$ of the HOTRG. 
It is expected that these two results are consistent at sufficiently large $l_{\rm max}$.

\begin{figure}[t]
 \centering
  \includegraphics[width=12cm]{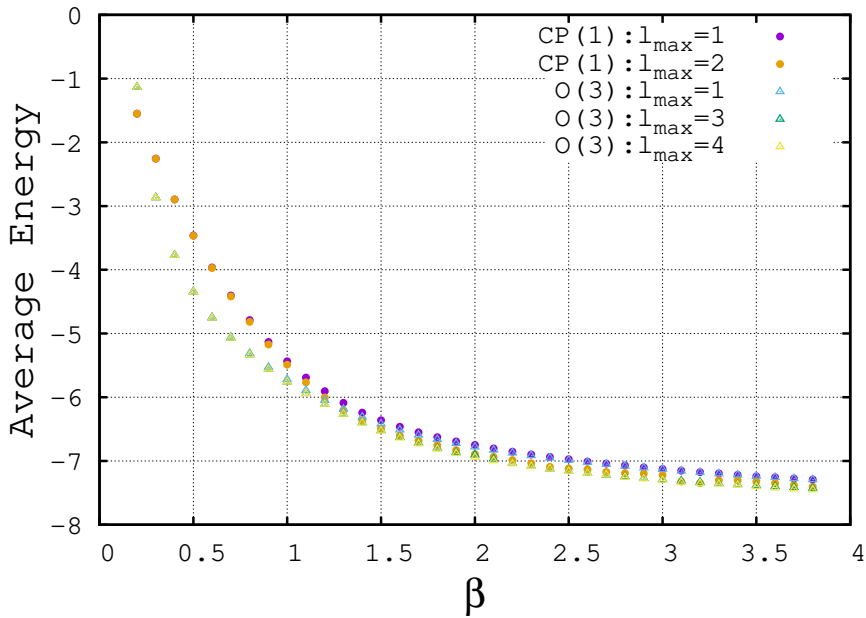}
 \caption{Average energy of CP(1) model and O(3) model computed by using HOTRG. The lattice size is $2^{20} \times 2^{20}$. The circle marks indicate the results of CP(1) model and the triangle marks indicate the results of  O(3) model.}
 \label{cp1o3}
\end{figure}

Next, Figure \ref{cp1o3} compares the result of the HOTRG with that of the O(3) nonlinear sigma model in two dimensions which is analyzed by the same method. Unmuth-Yockey {\it et al.} applied the HOTRG to the O(3) model \cite{Unmuth}. By following them, we compute the average energy of O(3) model. The energy of the two models is connected to each other in the continuum limit by the relation

\begin{align}
\frac{1}{\beta}+E_{\rm O(3)}(\beta)
=
E_{\rm CP(1)}(\beta)+6.
\end{align}
Using this relation, we mapped the result of the O(3) model into the graph. As $l_{\rm max}$ increases, the systematic errors decrease. In the limit $\beta=\infty$, these two results are expected to be consistent and in fact such a tendency is observed.

\section{Summary}

In this report, we show a tensor network representation of CP($N-1$) model without the $\theta$-term.
It is confirmed that the numerical results of CP(1) model at $\theta=0$ using the TRG method are consistent with that computed by Monte Carlo simulation and that of O(3) model which is analyzed by the same method in the region $\beta \gg 1$.

For our future work, we shall try to do implementation including the $\theta$-term. In the presence of this term, the integration over link valuables develops the additional terms, and furthermore $l$ no longer equals $m$. In this case, the computational cost of the TRG methods turns out to be very expensive and we may need some techniques to reduce the cost.

\end{document}